\documentclass[twoside,12pt]{article}
\usepackage{graphicx}
\usepackage{amssymb}

\begin{document}

\newcommand{\C}{{\mathbb C}}
\newcommand{\Z}{{\mathbb Z}}
\newcommand{\N}{{\mathbb N}}
\newcommand{\R}{{\mathbb R}}
\newcommand{\E}{{\mathbb E}}

\def\kasten{$~~\mbox{\hfil\vrule height6pt width5pt depth-1pt}$ }

\newcounter{roman}

\newtheorem{theorem}{Theorem}[section]

\par\noindent
{\Large \bf Remarks on some new Models of Interacting}
\par\noindent
{\Large \bf  Quantum Fields with Indefinite Metric }
\bigskip
\par\noindent
S. Albeverio${}^{1,2,3}$, H. Gottschalk${}^1$, and J.-L.
 Wu${}^{1,2,4}$
\medskip
\par\noindent
${}^1$ Fakult\"at und Institut f\"ur Mathematik der 
Ruhr-Universit\"at Bochum, D-44780 Bochum, Germany
\par\noindent
${}^2$ SFB237 Essen-Bochum-D\"usseldorf, Germany
\par\noindent
${}^3$ BiBoS Research Centre, Bielefeld-Bochum, Germany; and CERFIM,
Locarno, Switzerland
\par\noindent
${}^4$Institute of Applied Mathematics, 
Academia Sinica, Beijing 100080, P R China

\bigskip\par
\begin{abstract}{We study quantum field models in indefinite 
metric. We
introduce the modified Wightman axioms of Morchio and Strocchi as a
general framework of indefinite metric quantum field theory (QFT) 
and present concrete
interacting relativistic models obtained by analytical 
continuation from some
stochastic processes with Euclidean invariance. As a first step
towards scattering theory in indefinite metric QFT, we give a
proof of the spectral condition on the translation group for the relativistic
models.}
\end{abstract}

\section{Introduction}
\setcounter{section}{1}
By an argument due to Strocchi (for a review see e.g.\cite{S}) it became clear that in
quantized gauge theories which fulfill a so-called Gauss law (as e.g.
quantum electrodynamics) the postulate of locality of the fields is in
contradiction with the postulate of positivity of the inner product of
the Hilbert space associated to the quantum system. Technical reasons
(cf. \cite{S} p. 87) seem to indicate that it is better to drop
the latter condition and keep the former since the problems related to
the physical interpretation of a quantum system with indefinite inner product ("metric") can
be circumvented with the help of the formalism of Gupta-Bleuler gauge.

This motivated the study of Quantum Field Theories (QFT) in indefinite metric.
An axiomatic framework for indefinite metric QFT is given by the
modified Wightman axioms of Morchio and Strocchi \cite{MS}. Here the
requirement of positivity of the inner product of the Hilbert space is
replaced by the so-called Hilbert space structure condition, which
makes sure that the fields can still be realized as operators on a
Hilbert space. However, the Wightman functions of the theory are 
"vacuum expectation values" w.r.t. an inner product on the Hilbert
space which differs from the Hilbert space scalar product by a "metric operator".

In recent works \cite{AGW1,AGW2,AW} we studied interacting relativistic local quantum 
fields in the framework of indefinite metric QFT in any space-time dimension. 
We proved that these models fulfill the modified Wightman axioms. As
far as we know, these are the first models with these properties which
are constructed rigourously in a mathematical sense.

A natural question is that for a scattering theory for these models. This question 
has two aspects: $S$-matrix theory and Haag-Ruelle
theory. While for related vector models \cite{AGW2,AIK} we 
proved non-triviality of the associated scattering-(S)-matrix \cite{AGW4}, a similar
result is difficult to obtain for the scalar models considered here,
due to the conceptual difficulties of scattering theory for
quantum fields in which the one-particle states have a "smeared mass". 

On the other hand, this "smeared mass" in the case of the scalar models
smoothes the singularities of the Wightman functions, which turn out to be 
locally integrable and thus the Wightman functions are measures
(in contrast to this situation the Wightman functions in the vector case are 
first order distributional derivatives of measures).
This makes
it possible to construct unitary representations of the translation
group (but not of the entire Poincar\'e group!) for these models. These
theories thus have an energy-momentum operator, and we may ask for
the spectrum of this operator. We show, that it is located in the 
closed forward light cone.

Since the spectrum of the generator of the translation group and the
locality of the fields are the main indegrients of Haag-Ruelle theory in
the positive metric case, this might be a starting point for a Haag-Ruelle
theory in the indefinite metric case.

The article is organized as follows: In Section 2 we
briefly recall the axiomatic framework for indefinite metric QFT.
Section 3 sketches how to obtain Wightman functions including some
(non classical) interaction via analytic continuation 
from a corresponding 
Euclidean random field with some non Gaussian component. In the
last Section we give an outline how to verify the axioms of 
indefinite
metric QFT for these Wightman functions and sketch the proof of
the existence of unitary representations for  the translation group.
We also show, that the energy momentum operator has spectrum 
in the closed forward light cone. 

\section{The Axiomatic Framework for Indefinite Metric QFT}

Here we only treat the case of scalar chargeless fields 
(the generalization 
to higher spin and charged fields is 
straightforward). Let $s+1$ be the dimension of space-time. By ${\cal
S}_n$ we denote the space of complex Schwartz
functions on $\R^{(s+1)n}$ with the associated Schwartz topology 
and let
${\cal S}'_n$ be the topological dual spaces of tempered 
distributions.

The following \underline{modified Wightman axioms} for QFT in indefinite
metric are due to Morchio and Strocchi:

\

\noindent {\bf Axioms. A1)} \underline{Temperedness:} Let 
$\{ W_n \}_{n\in \N_0}$ be a
sequence of Wightman functions with $W_0=1$ and 
$W_n\in {\cal S}'_n$ for
$n\in\N$.

\noindent {\bf A2)} \underline{Poincar\'e invariance:} For any 
transformation $\Lambda$ in the proper, orthochronous Lorentz group
${\cal L}_+^\uparrow(\R\times\R^s)$ and any vector 
$a\in \R^{s+1}$ and
$n\in \N$
\begin{equation}
W_n(x_1,\ldots,x_n)=W_n(\Lambda^{-1}(x_1-a),\ldots,
\Lambda^{-1}(x_n-a))
\end{equation}
holds.

\noindent {\bf A3)} \underline{Spectral condition:} 
By the translation invariance (c.f. A2))
 we can define a tempered distribution $w_n\in {\cal S}_{n-1}'$ as
$w_n(y_1,\ldots,y_{n-1})=W_n(x_1,\ldots,x_n)$ with $y_j=x_j-x_{j+1}$. For
any $n\in \N$ the Fourier transform of $w_n$ has support in 
$(\bar V_0^{-
})^{\times (n-1)}$, where $\bar V_0^-$ (resp. $\bar V_0^+$) is the 
closed backward (resp. forward) 
light cone
in $\R\times \R^s$ (different conventions 
in some of the physical textbooks sometimes leads to the interchange 
of forward and backward
cones).

\noindent {\bf A4)} \underline{Locality:} For $n\geq 2$ and 
$x_j,x_{j+1}$ space-
like separated, we have the Bosonic commutation relations
\begin{equation}
W_n(x_1,\ldots,x_j,x_{j+1},\ldots,x_n)
=W_n(x_1,\ldots,x_{j+1},x_j,\ldots,x_n).
\end{equation}

\noindent {\bf A5)} \underline{Hermiticity:} For $n\in \N$ we have
\begin{equation}
W_n(x_1,\ldots,x_n)=\overline{W_n(x_n,\ldots,x_1)}.
\end{equation}

Up to this point the above axioms are the same as in standard QFT. 
In order to
get a indefinite metric QFT we have to drop the axiom 
of "positivity" of
the Wightman functions and replace it by the following:

\noindent {\bf A6)} \underline{Hilbert space structure condition:} 
For each $n\in \N$
there exists a Hilbert seminorm $p_n$ on ${\cal S}_n$ such that 
for any $f\in {\cal S}_j$
and $g\in {\cal S}_l$ the following inequality holds:
\begin{eqnarray}
|\int_{\R^{(s+1)(j+l)}}W_{j+l}(x_1,\ldots,x_{j+l})~
f(x_1,\ldots,x_j)&&\nonumber\\
\times ~g(x_{j+1},\ldots, g_{j+l})~dx_1\ldots dx_{j+l} | &\leq&
p_j(f)~p_l(g)~.
\end{eqnarray}
$\bullet$

\begin{theorem}[Morchio and Strocchi \cite{MS}]
Let $\{ W_n\}_{n\in \N_0}$ be a sequence of Wightman functions which
fulfill A1)-A6). Then there exist

\noindent (i) a Hilbert space ${\cal H}$ with
scalar product $(.,.)$, a distinguished vacuum vector $\Omega\in{\cal 
H}$ and an indefinite inner product $\langle.,.\rangle$ which
differs from $(.,.)$ only by a self adjoint 
\underline{metric operator}
$T$ with $T^2=1$, i.e. $\langle.,.\rangle=(.,T.)$;

\noindent (ii) a $T$-symmetric and local quantum field $\phi$, 
which is a distribution 
valued field operator $\phi(x)$ acting on a dense core
 ${\cal D}\subset{\cal H}$ with
$\phi(x)^*=T\phi(x)T$ and
\begin{equation}
[\phi(x),\phi(y)]_-=0 
\end{equation}
for $x$ and $y$ space-like separated. Furthermore $\phi$ is 
connected with the Wightman functions of the theory by
\begin{equation}
W_n(x_1,\ldots,x_n)=\langle \Omega,\phi(x_1)\cdots\phi(x_n)
\Omega\rangle~;
\end{equation}

\noindent (iii) a $T$-unitary representation $\cal U$ of the
connected orthochonous Poincar\'e group on ${\cal H}$, i.e. a
representation with $T{\cal U}^*T={\cal U}^{-1}$, s.t. $\Omega$ is
invariant under $\cal U$ and $\phi(x)$ transforms covariantly
\begin{equation}
{\cal U}(a,\Lambda)~\phi(x)~{\cal U}(a,\Lambda)^{-1}=\phi(\Lambda^{-
1}(x-a)).
\end{equation}
Furthermore, ${\cal U}$ fulfills the following spectral condition:
\begin{equation}
\int_{\R^{s+1}} \langle \Phi,{\cal U}(a,1)\Psi\rangle ~e^{iqa}~da =0~~
 \mbox{for
all } \Phi,\Psi \in {\cal D}
\end{equation}
if $q\not \in \bar V_0^-$. \kasten
\end{theorem}

A quadrupel $(({\cal H},(.,.),\Omega), T,
 \phi, {\cal U})$ is called a
\underline{QFT in indefinite metric}.

\section{From Euclidean Random Fields to Relativistic Wightman
Functions}

Since the work of Nelson \cite{N} it has become a paradigm in
QFT to construct relativistic QFT's by analytic continuation 
("reversing
the Wick rotation") from Euclidean random fields. In this picture, 
Gaussian
random fields correspond to non-interacting relativistic
 quantum fields.
Furthermore, random fields
corresponding to positive metric QFT fulfill some
 additional properties,
as e.g. the Markoff property. Dropping this requirement, we construct
non-Gaussian random fields which turn out to be the 
Euclidean analogue
of some indefinite metric QFT.

In this section we want to discuss heuristically the Euclidean random fields 
introduced in \cite{AGW1,AW} (related vector models are studied
 in \cite{AIK}).
 Let $\eta$ be a generalized
 white noise\cite{HKPS}, i.e. a ${\cal S}_1'$-valued
random variable which is independently and identically distributed in
different nonintersecting regions of the space-time $\R^{s+1}$ and 
has
an infinitely divisible (not necessarily Gaussian) probability law.
 Let $L$ be a continuously
invertible and symmetric (pseudo-) partial differential operator on
 ${\cal S}'$. We can then
solve the (pseudo-) stochastic partial differential equation
\begin{equation}
L\varphi(x)=\eta(x).
\end{equation}
Since $\eta(x)$ is by construction Euclidean invariant
(i.e. $\eta(x)=\eta(\Lambda^{-1}(x-a))$ in law for all 
$a\in \R^{s+1}$ and $\Lambda$
orthogonal), a necessary and
sufficient condition for $\varphi(x)$ beeing Euclidean invariant 
is the
Euclidean invariance of $L$. In the following we always assume
 this. Let
$G(x)$ denote the Green function of $L$ and let $\E$ be
 the expectation value w.r.t.
the probability space on which $\eta$ lives. By an explicit 
calculation one can then figure out 
the Schwinger (moment) functions $S_n\in {\cal S}_n'$ of $\varphi(x)$
\begin{eqnarray}
S_n(x_1,\ldots,x_n) &:=& \E \left[ \varphi(x_1)\cdots
\varphi(x_n)\right]
\nonumber \\
 &&\hspace{-.8in} = \sum_{I\in {\cal
P}^{(n)}} \prod_{\{j_1,\ldots,j_l\}\in I} 
\underbrace{c_l \int_{\R^{s+1}} G(x_{j_1}-
x)\cdots G(x_{j_l}-x) ~ dx~}_{=: S_l^T(x_{j_1},\ldots, x_{j_l})}.
\end{eqnarray}
Here the sum is over all partitions $I$ of $\{1,\ldots,n\}$ into
disjoint subsets and $c_l$ are constants which depend on the law of
$\eta(x)$. In particular, the law of $\eta(x)$ is Gaussian if and 
only
if $c_l=0$ for all $l>2$. The distributions $S_l^T$ are called the
truncated Schwinger functions of $\varphi (x)$. Thus, 
the higher order
$S_l^T,l>2,$ represent the interaction forces present in this model.

\

\noindent {\bf Remark.} In order to understand this interaction better,
 let us have a look at
the lattice analogue of $\varphi(x)$, i.e. we replace $\R ^{s+1}$ by
a bonded region $\Gamma$ in $\Z^{s+1}$. For all $x\in \Gamma$, 
$\eta(x)$ 
are indipendently identically distributed random variables with 
infinitely divisible law. Let $L_\Gamma$ be the lattice
analogue of $L$. Then the space of all possible 
configurations of the lattice field 
$\varphi(x)=L_\Gamma^{-1}\eta (x)$  is $\R^\Gamma$.
 Let $\rho $ denote the density of 
$\eta(x)$ w.r.t. the Lebesgue measure and let $c$ be the 
second moment of $\eta(x)$. 
For the expectation value of a bounded measurable 
function $C:\R^\Gamma \to \R$ one then gets the following path 
integral
representation \cite{AW}:
\begin{equation}
\E\left[ C\circ \varphi\right] = Z^{-1}\int _{\R^\Gamma} C(\varphi) e^{-
\int_{\R^{s+1}}
c (\varphi(x)L_\Gamma^2 \varphi(x)) - V(L_\Gamma \varphi(x)) 
d_\Gamma x}
{\cal D}_\Gamma \varphi .
\end{equation}
Here ${\cal D}_\Gamma\varphi$ is the lattice path integral measure, $Z$
is a normalization constant,
$d_\Gamma x=\sum_{x'\in \Gamma}\delta(x-x')dx$ is the lattice 
Lebesgue
measure and $V(t):=\log\rho (t)+ct^2$ is 
the "potential function". 

Obviously, for $L=(-\bigtriangleup +m^2)^{1\over 2}$ 
the quadratic term in the above
action is the usual kinetic energy term of the free field 
and thus $V$ 
is responsible for the interaction. For $\eta(x)$ to be 
Gaussian we clearly get $V\equiv 0$ and
there is no interaction. But if $\eta (x)$ contains a non Gaussian
component we have $V\not \equiv 0$. Nevertheless, the potential
 term $
V(L_\Gamma\varphi(x))$ is different from the usual potential
 term e.g.
in  $P(\varphi)_2$-theory which is of the form $\tilde V (\varphi
(x))$. Since $L$ in general is a nonlocal operator, the 
interaction in our
model exhibits a kind of "non locality" which has to be made
 responsible
for the lack of the Markoff property of the field 
$\varphi (x)$ \cite{AGW1}. 
On the other hand, this "non locality" does not destroy the 
relativistic
requirement of "locality", since, roughly speaking, in the 
Euclidean region 
there are only space-like distances and thus the non-local
 smearing in the
fields due to $L$ only smears space-like separated amplitudes
 of the field.
On the other hand, this "smearing" might be responsible
 for the analytic
well behaviour of these models, since "smearing" implies some
 regularization
of the distributional fields $\varphi(x)$. However, a precise
mathematical formulation for these "explanations" has 
not been obtained
yet. $\bullet$

Let us explain now how to do the analytical continuation from purely
imaginary Euclidean time back to purely real relativistic time. 
This can
be done by representing the truncated Schwinger functions
$S_n^T(x_1,\ldots,x_n)$ as Laplace transform of the Fourier transform
$\hat W_n^T(x_1,\ldots,x_n)$ of the truncated Wightman functions 
$W_n^T$
of the model, where all $W_n^T, n\in \N,$ fulfill the
 spectral condition:
\begin{equation}
S^{T}_{n}(x_1,\cdots,x_n)={ (2\pi)^{\scriptscriptstyle 
-\frac{(s+1)n}{2}}}
\! \! \int_{{\R}^{(s+1)n}}\hspace{-.2in}e^{\sum^n_{l=1}
ik^0_l(ix^0_l)+i\vec{k}_{l}\vec{x}_{l}}
\hat{W}^T_n(k_1,\cdots,k_n)~dk_1\cdots dk_n,
\end{equation}
for $x_1^0<x_2^0<\cdots<x_n^0$. To get $W_n^T$ one then just has to
replace on the right hand side $ix_l^0$ by $x_l^0$.

Let us take the particular case where 
$L=(-\bigtriangleup +m^2)^\alpha$ for
$0<\alpha \leq 1/2$. By a lenghtly but explicit calculation
 we proved in
\cite{AGW1} that such a representation indeed does exist. 
The formulae
for the $\hat W_n^T, n>1,$ are:

\begin{equation}
\hat{W}^T_{n}(k_1,\cdots,k_n) := 
c_n'\left\{\sum^n_{j=1}\prod^{j-1}_{l=1}
\mu^-_\alpha(k_l)\mu_\alpha(k_j)
\prod^n_{l=j+1}\mu^+_\alpha(k_l)\right\}\delta(\sum^n_{l=1}k_l).
\end{equation}
with $c_n'=c_n2^{n-1}(2\pi)^{s+1}$,
\begin{eqnarray}
\mu^+_\alpha(k) &:=& (2\pi)^{-{(s+1)\over2}}\sin\pi
\alpha 1_{\{k^2>m^2,k^0>0\}}(k)
                   \frac{1}{(k^2-m^2)^\alpha}  \nonumber \\
\mu^-_\alpha(k) &:=& (2\pi)^{-{(s+1)\over2}}\sin\pi
\alpha 1_{\{k^2>m^2,k^0<0\}}(k)
                   \frac{1}{(k^2-m^2)^\alpha} \nonumber \\
\mu_\alpha(k) &:=& (2\pi)^{-{(s+1)\over2}}\!\left(\cos\pi
\alpha 1_{\{k^2>m^2\}}(k)+
                   1_{\{k^2<m^2\}}(k)\right)\frac{1}{|k^2-m^2|^\alpha}
\end{eqnarray}  
for $n\geq 3$ as well as $n=2,\alpha<1/2$ and in the case 
$n=2,\alpha =
1/2$
\begin{equation}
\hat W_2^T=2\pi c_2 1_{\{
k_1^0<0\}}(k_1)\delta (k^2-m^2) \delta(k_1+k_2)~.
\end{equation}
Here we have used the notation $k^2={k^0}^2-|\vec
k|^2$.

Let the Wightman functions $W_n$ be defined by
 the truncated Wightman
 functions $W_l^T$ in the way of formula (10). Then the 
Wightman functions
are given as the analytic continuation of the
 Schwinger functions of the
model. The spectral property of the $W_n$ can be
checked by hand using the Eqn. (13)-(15) and the Schwinger functions 
by definition are real, symmetric and Euclidean invariant. 
The following
theorem therefore follows from the Osterwalder -- Schrader 
reconstruction theorem
\cite{OS}:

\begin{theorem}[Main Theorem of \cite{AGW1}]
The sequence of Wightman functions $\{ W_n\}_{n\in \N_0}$
 fulfills the
axioms A1)-A5). \kasten
\end{theorem}

\section{Hilbert Space Structures with Unitary Translation Group}

In order to prove that the Wightman functions $\{ W_n\}_{n\in \N_0}$
belong to an indefinite metric QFT, it remains to verify A6). 
This can
be done as follows. First, for $n\in \N$, we choose a special 
system of Schwartz norms
\begin{equation}
\|f\|_{K,N}:=\sup_{\stackrel{\scriptstyle x_1,\ldots ,x_n\in
 \R^{s+1}}
{ 0\leq|\beta_1|,\ldots,|\beta_n|\leq K}}\left|
\left[ \prod_{l=1}^n(1+|x_l|^2)^{N/2}\left( {\partial
\over \partial x_l}\right)^{\beta_l} \right]f(x_1,\ldots,x_n)
\right|~.
\end{equation}
on ${\cal S}_n$. Based on a kind of "quantitative nuclear
 theorem" related to this
system of norms, it is possible to prove the following sufficient
 \underline{Hilbert
space structure} \underline{condition for the
 truncated Wightman functions}:

\begin{theorem}[\cite{AGW2}]
Let $K,N\in \N_0$ be fixed. If for all $n\in \N$ $\hat W_n^T$
 are continuous
with respect to $\|.\|_{K,N}$, then the sequence of Wightman 
functions
\linebreak $\{ W_n\}_{n\in \N_0}$ fulfills A6). \kasten
\end{theorem}

For a generalization of this theorem see \cite{H}.

By explicit calculation using the formulae (13)-(15) we proved in
\cite{AGW2} the followig inequalities: For $n\in \N$ there exist 
constants $a_n>0$ s.t.
\begin{equation}
| \int_{\R^{(s+1)n}}\hat W^T_n(k_1,\ldots,k_n)~f(k_1,\ldots,k_n) ~
dk_1\cdots dk_n | \leq a_n \| f\|_{0,2s+2}
\end{equation}
holds for all $f\in {\cal S}_n$. Now from the Theorems 2.1, 
3.1 and 4.1 
we immediately get

\begin{theorem}[Main Theorem of \cite{AGW2}]
 The Wightman functions
$\{W_n\}_{n\in \N_0}$ constructed in Section 3 fulfill all modified 
Wightman axioms A1)-A6). Thus, there exists a indefinite 
metric QFT with
Wightman functions $\{W_n\}_{n\in \N_0}$.  \kasten
\end{theorem}

For a given sequence of Wightman functions there can be different
Hilbert space structures, which are, however, determined 
by the Hilbert
seminorms on the right hand side of Eq. (4). In particular,
 if these
seminorms can be chosen to be translation invariant, the 
metric operator $T$
is also translation invariant, i.e. $[{\cal U}(a,1),T]_-=0$ for all
$a\in \R^{s+1}$. This implies that at least the translations are
represented by unitary operators. In this case it makes sense to 
speak of the
generator $P$ of the translation group, which is a self 
adjoint vector
operator on $\cal H$ with spectrum $\mbox{spec}(P)$. The following
theorem is a new result on the existence of such Hilbert space structures.
For a detailed proof we refer to \cite{AGW3}.

\begin{theorem}
For the given sequence of Wightman functions 
$\{W_n\}_{n\in \N_0}$ the
Hilbert seminorms in the Hilbert space structure condition Eq. (4) 
can be chosen translation invariant. 
Thus, there is a Hilbert space structure s.t. the 
representation of the
translation group ${\cal U}(a,1)$ is unitary. If $P$ denotes the
generator of ${\cal U}(a,1)$, then $\mbox{\rm spec}(P)\subseteq \bar
V^+_0$.
\end{theorem}

\noindent {\bf Proof.} First let us assume that Eq. (4) is fulfilled
with $p_j, j\in\N,$ translation invariant. Then the positive
semidefinite inner product on ${\cal D}$ constructed in \cite{MS} is
translation invariant and thus the obtained representation of the
translations ${\cal U}(a,1)$ on the Hilbert space related to this inner
product and the metric operator $T'$ commute. However, in general $T'$
is not continuously invertible and thus does not fulfill ${T'}^2=1$. But
from the given metric operator one can pass to a new metric operator
with $T^2=1$ using the procedure described in \cite{MS} (see also
\cite{AGW2}). But the steps of this procedure consist of changing the
inner product two times by a function of the metric operator.
 Since ${\cal U}$ commutes with such functions (since it commutes with $T'$),
we get that ${\cal U}$ commutes with the resulting metric
operator $T$. Thus we get a unitary representation of the translation
group.

Let $P$ be the generator of this representation. In order to prove
$\mbox{spec}(P)\subseteq \bar V_0^+$ we first write Eq. (8) in the
equivalent form
\begin{equation}
\int_{\R^{s+1}} \overline{(\Phi,{\cal U}(a,1)\Psi)}\hat f(a) da=0
\end{equation}
for all $\Phi\in T{\cal D}, \Psi \in {\cal D}$ and $f\in {\cal S}_1$
with $\mbox{supp} f \cap \bar V_0^+=\emptyset$. Since
$T{\cal D}$ and ${\cal D}$ are dense in ${\cal H}$ we have sequences
$\Phi_n \in T{\cal D}$ and $\Psi_n \in {\cal D}$ with $\Phi_n\to \Phi$ and
$\Psi_n\to \Psi$ in ${\cal H}$. Thus
\begin{eqnarray}
&&\left| \int_{\R^{s+1}} \overline{(\Phi,{\cal U}(a,1)\Psi)}\hat f(a)
da\right| \nonumber \\
&\leq&\sup_{a\in \R^{s+1}}|(\Phi,{\cal U}(a,1)\Psi)-
(\Phi_n,{\cal U}(a,1)\Psi_n)|\int_{\R^{s+1}} |\hat f(a)| da\nonumber \\
&\leq& (\sup_{n\in \N}\{\|\Phi_n\|\} \|\Psi-\Psi_n\|+\|\Psi\| \|\Phi-\Phi_n\|)\|\hat
f\|_{L^1(\R^{s+1},dx)}\rightarrow 0
\end{eqnarray}
as $n\to \infty$.
Thus Eq. (18) holds for all $\Phi, \Psi
\in {\cal H}$. Let $E$ denote the spectral measure of $P$. Then the left
hand side of Eq. (18) is equal to 
\begin{equation}
(2\pi)^{-
(s+1)/2}\int_{\R^{s+1}} f(\lambda) d(\Phi, E(\lambda) \Psi)~,
\end{equation} 
which shows
that $\mbox{spec}(P)=\mbox{supp}(E)\subseteq \bar V_0^+$.

It remains to show the existence of translation invariant Hilbert
seminorms $p_j,j\in\N,$ s.t. Eq. (4) holds. Note that from the formulae
(10) and (13)-(15) it follows that $\hat W_n$ is a positive measure for
$n\in\N$. From Parseval theorem, the Cauchy Schwarz inequality and
Fubini theorem we therefore get that the left hand side of Eq. (4) is
smaller or equal than
\begin{eqnarray}
&&\left( \int_{\R^{(s+1)j}}|\hat f(k_1,\ldots,k_j)|^2\left[
\prod_{r=1}^j(1+|k_r|^2)^{s+1}\int_{\R^{(s+1)l}} \hat
W_{j+l}(k_1,\ldots,k_{j+l})\right. \right.\nonumber\\
&&\left. \left. \times\prod_{r=j+1}^{j+l}(1+|k_r|^2)^{-(s+1)} ~
dk_{j+1}\cdots dk_{j+l}\right] dk_1\cdots dk_j \right)^{1/2}\nonumber
\\
&&\times\left( \int_{\R^{(s+1)l}}|\hat g(k_{j+1},\ldots,k_{j+l})|^2\left[
\prod_{r=j+1}^{j+l}(1+|k_r|^2)^{s+1}\right. \right.\nonumber\\
&& \times\int_{\R^{(s+1)j}} \hat
W_{j+l}(k_1,\ldots,k_{j+l})\nonumber \\
&&\left. \left.\times \prod_{r=1}^{j}(1+|k_r|^2)^{-(s+1)} ~
dk_{1}\cdots dk_{j}\right] dk_{j+1}\cdots dk_{j+l}
\right)^{1/2}
\end{eqnarray}
If we can now show that for $j,l \in \N$ there exists a measure $M_j$
on $\R^{(s+1)j}$ and constants $C_{j,l}>0$ such that
\begin{equation}
\int_{\R^{(s+1)j}}|\hat f|^2[\ldots ]dk_1\cdots dk_j\leq
C_{j,l}\int_{\R^{(s+1)j}}|\hat f|^2dM_j < \infty
\end{equation}
we have finished the proof, since we may define
\begin{equation}
p_j(f)= (\max_{1\leq r,q\leq j}C_{r,q}+1) \left( \int_{\R^{(s+1)j}}|
\hat f|^2dM_j\right)^{1/2},
\end{equation}
and from (21)-(23) we easily get Eq. (4).
$p_j$ obviously is translation invariant, since the translation group
acts on $\hat f$ by multiplication with $\exp\{i\sum_{r=1}^jk_r\cdot a\}$
wich does not change $|\hat f|^2$ and hence does not change $p_j(f)$.

The proof of the existence of the measures $M_j$ is technical and not
very instructive. Here we only give formulae for suitable measures $M_j$
for the case $\alpha < 1/2$ which is a little easier to treat than the
case $\alpha = 1/2$. Using estimates similar to those in
Subsection 4.1 of \cite{AGW2} one can prove that the measures
\begin{eqnarray}
dM_j(k_1,\ldots,k_j) &=& \sum_{I\in {\cal P}^{(j)}}\prod
_{\{j_1,\ldots j_l\}\in I}\prod_{r=1}^l   (1+|k_{j_r}|^2)^{s+1} \nonumber \\ 
&&\hspace{-1.5in}\times\left[\delta_{l,1}{1\over|k_{j_1}^2-
m^2|^{2\alpha}}+ \delta_{l,2}{1\over|k_{j_1}^2-
m^2|^{2\alpha}}\delta(k_{j_1}+k_{j_2})+(1-\delta_{l,2})\prod_{r=1}^l 	{1\over |
k_{j_r}^2-m^2|^{\alpha}}\right.\nonumber \\
&&\left.\hspace{-.5in}+(1-\delta_{l,1}-
\delta_{l,2}){|(\sum_{r=1}^lk_{j_r})^2-m^2|^{-
\alpha}\over
\prod_{r=1}^l|k_{j_r}^2-m^2|^\alpha}\right]dk_1\ldots dk_j
\end{eqnarray}
fulfill the requirements of Eq. (23). (Note that by $2\alpha < 1$
$M_j$ has a locally integrable density w.r.t. the Lebesgue measure
which implies that the second estimate in (22) holds.)\kasten

Since the spectrum of $P$ together with the locality of the fields
in the positive metric case are the main properties on which 
Haag-Ruelle theory (i.e. the proof of 
existence of asymptotic "scattering" states) is based 
\cite{RS}, we consider Theorem 4.3 as a first step towards 
an analogous theory in the indefinite metric case. 
However, there are
other properties of indefinite metric QFT which differ significantly
from the case of positive metric. E.g. the uniqueness of the vacuum
cannot hold in case $T$ commutes with ${\cal U}$, since 
$T\Omega\not \in \C \Omega$ is translation invariant, too. Therefore,
the operation of truncation has to be carried out w.r.t. 
the projection
onto the $0$-eigenspace of $P$ instead of the projection onto $\C\Omega$. 
But then the truncated objects are operator valued functions acting on the 
$0$-eigenspace rather
than distributions.
Possibly, these operators on a space of at least two dimensions
do not commute, which could destroy 
"locality" of the truncated objects. But that property plays a 
crucial r\^ole in
Haag Ruelle theory\cite{RS}. We will investigate
 these problems in \cite{AGW3}.

\

\noindent{\bf Acknowledgements.} One of us (H.G.) would like to thank 
 the E.C. science organization for financial support
through the project "Stochastic dynamics and statistical mechanics 
and interfaces", making it possible for him to attend the 29th
 Symposium on Mathematical
Physics in Thoru\'n. The financial support of D.F.G. is gratefully
acknowlegded.

\end{document}